# Ultra-rapid broadband mid-infrared spectral tuning and sensing

XIAOSHUAI MA,[1,†] TIANJIAN LV, [1,†] XUDONG ZHU, [1] MING YAN,[1,2,*] AND HEPING ZENG[1,2,3]

[1]*State Key Laboratory of Precision Spectroscopy, and Hainan Institute, East China Normal University, Shanghai 200062, China*
[2]*Chongqing Key Laboratory of Precision Optics, Chongqing Institute of East China Normal University, Chongqing 401120, China*
[3]*Jinan Institute of Quantum Technology, Jinan, Shandong 250101, China*
[†]*These authors contributed equally.*
*\*myan@lps.ecnu.edu.cn*

**Abstract:** Tunable mid-infrared lasers are essential for optical sensing and imaging. Existing technologies, however, face challenges in simultaneously achieving broadband spectral tunability and ultra-rapid scan rates, limiting their utility in dynamic scenarios such as real-time characterization of multiple molecular absorption bands. Here, we present a high-speed approach for broadband wavelength sweeping in the mid-infrared region, leveraging spectral focusing via difference frequency generation between a chirped fiber laser and an asynchronous, frequency-modulated electro-optic comb. This method enables pulse-to-pulse spectral tuning at a speed of 5.6 THz/µs with 380 elements. Applied to spectroscopic sensing, our technique achieves broad spectral coverage (2600–3780 $cm^{-1}$) with moderate spectral resolution (8 $cm^{-1}$) and rapid acquisition times (~6.3 µs). Notably, the controllable electro-optic comb facilitates high scan rates of up to 2 Mscans/s across the full spectral range (corresponding to a speed of 60 THz/µs), with trade-offs in number of elements (~30) and spectral point spacing or resolution (33 $cm^{-1}$). Nevertheless, these capabilities make our platform highly promising for applications such as flow cytometry, chemical reaction monitoring, and mid-IR ranging and imaging.

## 1. Introduction

Wavelength-tunable or frequency-swept lasers have proven instrumental across a wide scope of applications, including spectroscopic sensing [1], biological imaging [2], and light detection and ranging (LIDAR) [3]. In spectroscopy, combining a tunable laser with a photodetector creates a simple yet highly versatile instrument, which obviates the need for dispersive optics and sensor arrays in spectral analysis, delivering unmatched spectral resolution compared to traditional dispersive spectrometers. Notably, frequency-swept spectroscopy surpasses Fourier-transform spectroscopy in both signal-to-noise ratio (SNR) and measurement accuracy. Fourier-transform techniques suffer from penalties associated with multiplexed detection [4] and instrumental line shape distortions arising from the mechanical constraints of a moving arm [5]. The advent of optical frequency combs and dual-comb spectroscopy [5–10] migrates some of these problems by enabling motionless multi-heterodyne detection using two frequency combs, but remains constrained by limitations in SNR and sensitivity [4]. In the realm of LIDAR, integrating a tunable laser with dispersive optics for spectral-spatial encoding enables inertia-free parallel distance metrology [3] and facilitates multi-dimensional imaging [11]. These advancements underscore the pivotal role of tunable lasers in driving progress across diverse scientific and technological domains [12].

For many of these applications, accessing the mid-infrared (mid-IR) regime (typically 2–20 µm) is crucial. In spectroscopy, molecular fundamental ro-vibrational transitions are prominent in the mid-IR, enabling strong light absorption and enhanced sensitivity for chemical sensing [13]. In LIDAR, the mid-IR spectrum encompasses two critical atmospheric transmission windows (3–5 µm and 8–12 µm) [14, 15], which support long-range applications with reduced attenuation. For optical imaging, mid-IR light can penetrate materials opaque to visible or near-IR wavelengths, such as plastics, making it invaluable for imaging through barriers or within subsurface structures [16]. Despite these advantages, mid-IR tunable sources remain less developed compared to their visible/near-IR counterparts. Challenges such as limited tuning speed and constrained spectral coverage hinder their broader adoption in these applications.

As the workhorses of this domain, quantum cascade lasers (QCLs) are available across an exceptionally broad spectral range, spanning the mid-IR to terahertz regimes [17, 18]. However, individual QCLs cover only a narrow portion of the spectrum and are limited in tuning speed or scan rates (typically 10s of kHz) [19, 20]. Moreover, while most QCLs operate in continuous-wave mode—achieving tremendous success in frequency-swept spectroscopy—they face challenges in time-resolved or nonlinear measurements [21]. Recent advancements in QCL-based frequency comb technology show great promise in addressing some of these limitations [22, 23]. Alternatively, nonlinear optical techniques such as optical parametric oscillation (OPO) [24, 25] and difference frequency generation (DFG) [26-28] can produce mid-IR light with a broad tunable range in both continuous-wave and pulsed modes. However, neither mechanical adjustments (e.g., changing the OPO cavity length or the incident angle) nor stationary measures (e.g., tuning pump wavelengths via laser diode current or using acousto-optic tunable filters [16]) are sufficiently time-efficient. As a result, ultra-rapid, broadband spectral tuning remains challenging, particularly in the critical mid-IR region.

In this paper, we introduce a spectral focusing DFG scheme to address this challenge. Spectral focusing, commonly employed to enhance the spectral resolution in coherent Raman spectroscopy [29, 30], is adapted here for the first time—coupled with controllable asynchronous pulses—for mid-IR light generation. Innovatively, this approach allows fast spectral tuning on a pulse-to-pulse basis. In particular, our approach generates a time-domain sequence of mid-IR pulses, each with a specific center wavelength (or "color") and a narrow bandwidth (~8 cm$^{-1}$), spanning a spectral range from 2600 to 3780 cm$^{-1}$. Experimentally, our setup comprises a broadband mode-locked fiber laser and a frequency-modulated electro-optical (EO) laser comb. These lasers operate with slightly different repetition frequencies, akin to a dual-comb system. A critical feature of our scheme is the modulation of the EO comb's repetition frequency, enabling effective control of the asynchronous time delay. This modulation facilitates rapid spectral tuning at a scan rate of up to 2 Mscans/s. From an application standpoint, the mid-IR source supports broadband spectral measurements (>1000 cm$^{-1}$) within a few microseconds, or even faster. We demonstrate its capabilities through the analysis of chemical liquids, showcasing its potential for high-speed, high-resolution spectroscopic applications.

## 2. Basic principles

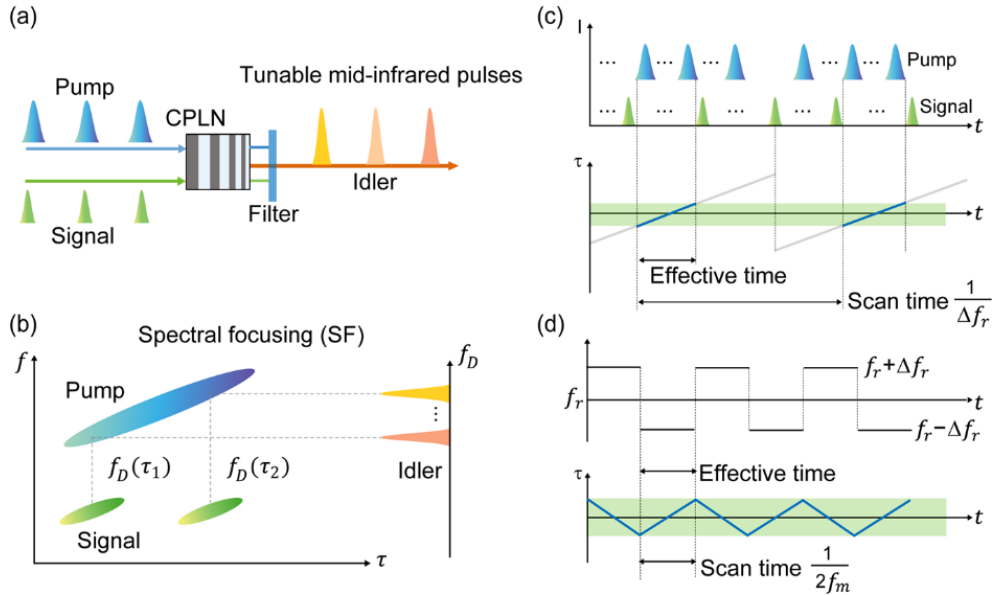

Fig. 1. Basic principles. Schematics of (a) difference frequency generation, (b) spectral focusing, and (c) optical sampling with asynchronous pulses. (d) Modulating the repetition frequency of signal pulses to improve the scan rate. CPLN, chirped-poling lithium niobate; $t$, measurement time; $\tau$, the relative time delay between the pump and signal pulses; $f_r$, the repetition frequency of the pump laser; $\Delta f_r$, the repetition frequency difference; $f_D$, the center frequency of the generated idler beam; $f_m$, the modulation frequency.

In our approach (Fig. 1(a)), mid-IR pulses are generated via difference-frequency generation (DFG) between two trains of chirp-matched pulses—referred to as the pump and signal pulses—in a chirped-poling lithium niobate (CPLN) crystal, which supports broadband frequency conversion. Notably, our method integrates two key techniques.

First, a spectral focusing technique is employed to narrow the spectral bandwidth of the mid-IR pulses [31, 32], thereby defining the resolution limit for spectroscopic applications. The principle of spectral focusing is illustrated in Fig. 1(b). For a linearly chirped pulse, the instantaneous central frequency $f(t)$ can be expressed as

$$f(t) = f_c + \beta \cdot t,$$

where the chirp parameter $\beta$ describes the linear slope of the central frequency ($f_c$), and $t$ is the time. In DFG, a mid-infrared pulse is generated at the instantaneous frequency difference, $f_D(\tau)$, which linearly changes with the relative time delay ($\tau$) between the pump and the signal pulses, as

$$f_D(\tau) = \Omega_0 + \beta \cdot \tau,$$

where $\Omega_0$ is the pump and signal center frequency difference. By selecting signal and pump lasers spectrally centered at approximately 1.03 μm (280 THz) and 1.55 μm (192 THz), respectively, we generate idler pulses in the 3-μm region ($\Omega_0 \sim 90$ THz). The bandwidth of each pulse is determined by the degree of chirp matching between the pump and signal pulses.

Secondly, an asynchronous time-delay scheme is utilized to tune the idler frequency ($f_D$) in an inertia-free manner [33]. As shown in Fig. 1(c), the pump and signal lasers have slightly different repetition frequencies ($f_r$ and $f_r + \Delta f_r$, respectively), causing their relative delay ($\tau$) to change linearly with a step size of $\Delta f_r / f_r^2$. However, this scheme results in a poor duty cycle, meaning the effective time window during which the two pulses temporally overlap is much smaller than the scan period ($T = 1/\Delta f_r$), as shown in Fig. 1(c). The primary reason is that the pulse durations of the two lasers are much shorter than their repetition periods. This leads to significant dwell time and limits the refresh rate of spectral measurements [34, 35].

To address this issue, we modulate the signal's repetition frequency between $f_r - \Delta f_r$ and $f_r + \Delta f_r$ at a modulation frequency of $f_m$, as shown in Fig. 1(d). Consequently, we can scan $\tau$ within the pulses' overlapping region, achieving a scan time equal to the effective time, or a 100% duty cycle. This approach, which has proven successful in dual-comb spectroscopy [29, 30], has not yet been explored for boosting the scan rate of a tunable source. An additional advantage of our scheme lies in the use of a cavity-free EO comb, whose repetition frequency can be rapidly modulated by a radio-frequency synthesizer at speeds far exceeding those of mode-locked lasers [29].

## 3. Results

### 3.1 Mid-infrared color pulse generation and characterization

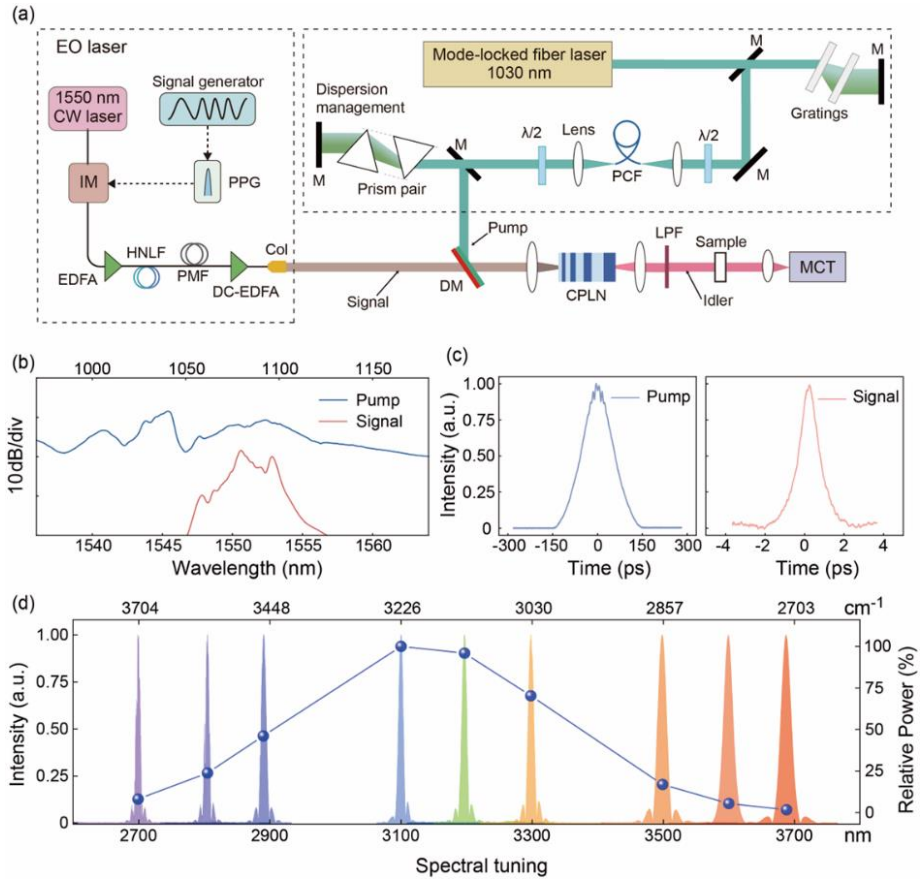

Fig. 2. Generation of broadband tunable mid-infrared light. (a) Experimental setup. CW, continuous-wave laser; IM, intensity modulator; PPG, picosecond pulse generator; DC-EDFA, double-clad Er-doped fiber amplifier; HNLF, highly nonlinear fiber; SMF, single-mode fiber; YDFA, Yb-doped fiber amplifier; Col, fiber collimator; M, mirror; λ/2, half-wave plate; PCF, photonic crystal fiber; DM, dichroic mirror; CPLN, chirped-poling lithium niobate crystal; LPF, long-pass filter; MCT, HgCdTe photodetector. (b) Spectral characterization of the pump and the signal light. (c) Measured autocorrelation traces of the pump and the signal pulses. (d) Spectra of the narrowband mid-infrared light measured with a commercial spectrometer. The spectral intensity is normalized, and the blue lines depict the relative intensities of different spectral components.

As shown in Fig. 2a, our mid-IR source utilizes a broadband fiber laser as the pump light and an EO laser as the signal light. The pump light originates from a mode-locked Yb-doped fiber oscillator at 1.03 μm. With the help of an intracavity piezoelectric actuator and a servo system, the repetition rate is stabilized at 60.5 MHz with a standard deviation less than 3 mHz. This laser contains an ytterbium-doped fiber amplifier that boosts the average power to approximately 1.6 W. The output pulses are compressed to ~120 fs using a couple of transmission gratings (1250 lines/mm). These pulses are then focused into a 10-cm-long photonic crystal fiber (PCF; SC-3.7-975, NKT Photonics) using an aspheric lens (C330TMD-B, Thorlabs, f = 3.1 mm). This setup generates a spectral range exceeding 1300 cm$^{-1}$ (from 1000 to 1150 nm) with an average power of >200 mW, as shown in Fig. 2(b). A pair of prisms are used for precise chirp control.

In the EO laser, a narrow-linewidth continuous-wave (CW) laser (linewidth <10 kHz, 100 mW; ID Photonics) is sent into an intensity modulator (IM) with a 40-GHz bandwidth (KY-MU-15-DQ-A, Keyang Photonics). The IM is driven by a pulse generator (LaseGen) that produces 30-ps electrical pulses, triggered by a radio-frequency signal generator (DSG3060, Rigol). As a result, the IM outputs 30-ps optical pulses. After amplification by a custom-built erbium-doped fiber amplifier (EDFA), the optical power is increased to approximately 350 mW. The output undergoes spectral broadening in a segment of highly nonlinear fiber (HNLF) with a linear dispersion slope of 0.03 ps/(nm²·km) and a nonlinear coefficient of about 10 W$^{-1}$·km$^{-1}$. This is followed by dispersion compensation using a long polarization-maintaining fiber. Finally, the EO laser is further amplified to >0.5 W using a double-clad EDFA.

The net group velocity dispersion for each laser is carefully managed to ensure that the pump and signal pulses are linearly dispersed with matched chirps before being used for DFG. For a highly chirped Gaussian pulse, the linear chirp parameter $\beta$ can be estimated as $\beta \approx \Delta f / \Delta \tau$, where $\Delta f$ is the spectral full width at half maximum (FWHM) and $\Delta \tau$ is the pulse duration. As shown in Fig. 2(b), the pump spectral FWHM is approximately 11 THz, and the signal FWHM is about 0.09 THz. The measured autocorrelation traces of the pump and signal pulses, shown in Fig. 2(c), indicate a pump pulse duration of ~90 ps and a signal pulse duration of ~750 fs. Consequently, $\beta$ is estimated to be ~0.12 THz/ps for both lasers.

To produce mid-IR pulses, the chirp-matched pump and signal beams are spatially overlapped using a dichroic mirror (DMLP1180, Thorlabs), and then focused into a 2-cm-long chirped-poling lithium niobate (CPLN) crystal (Castech) for DFG. The CPLN crystal features a linearly increasing poling period (from 23 to 31 μm), facilitating adiabatic quasi-phase matching across a broad spectral range. The idler beam is collimated and filtered through a long-pass filter (cutoff wavelength at 2.6 μm), which blocks the fundamental beams.

We confirm the mid-IR spectral tuning range by operating the pump and signal lasers in synchronized mode. By adjusting their relative pulse delay, we measure the mid-IR spectra using a standard Fourier-transform infrared spectrometer (FTIR, resolution of 0.1 cm$^{-1}$, PerkinElmer). Figure 2(d) exemplifies several recorded spectra, spanning from 2600 to 3700 nm (or 2703 to 3704 cm$^{-1}$). The FWHM of each tuning spectrum is ~8 cm$^{-1}$, indicating the spectral width of a mid-IR color pulse achieved in asynchronous mode. This also sets a resolution limit for molecular spectroscopy using our source. Also, the maximum output power of 1 mW is achieved at 3100 cm$^{-1}$ (blue dots in Fig. 2(d)), corresponding to a maximum energy per spectral element of 16.5 pJ.

*3.2 Ultra-rapid, broadband spectral measurements*

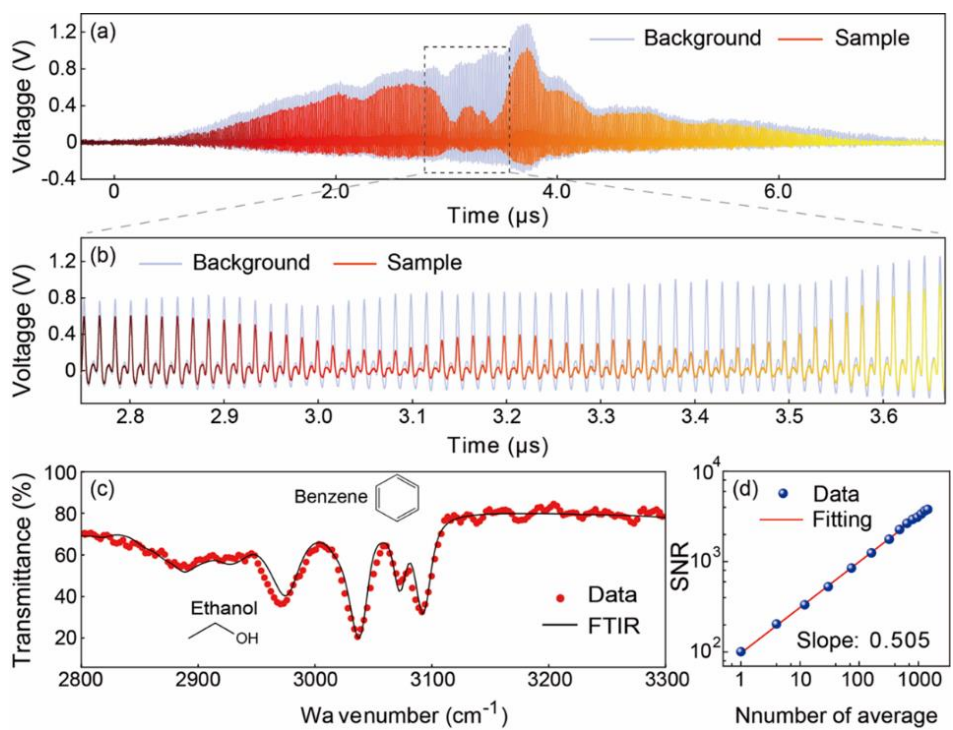

Fig. 3. Results of temporal measurements and spectral reconstruction. (a) Pulse trains recorded in the time domain. (b) Enlarged view of the recoded pulses. The pulses separated by 16.5 ns, determined by the pump laser's repetition frequency. (c) The normalized spectrum reconstructed from a single measurement and its comparison with the reference trace measured by a commercial spectrometer. (d) Signal-to-noise ratio (SNR) versus number of averages.

We then demonstrate the capability of our source for broadband molecular fingerprinting. To this end, the mid-IR light is directed into a thin cuvette filled with chemical liquids and subsequently detected by a single mid-infrared detector (bandwidth of 250 MHz, Vigo). The detector output is recorded by a digital oscilloscope (bandwidth of 1 GHz, LeCroy) with a sample rate of 500 MS/s. Figure 3(a) shows two mid-IR pulse trains recorded with and without a liquid chemical sample (a mixture of benzene and ethanol in a 2:1 concentration ratio). For these measurements, the repetition frequency of the EO laser is set to 302.5 MHz + 14 kHz, close to the fifth harmonic of the fiber laser's repetition frequency. The EO laser's overall performance is optimal at this frequency. Consequently, each recorded time-domain trace lasts approximately 6.3 μs, corresponding to a full spectral coverage of 1180 cm$^{-1}$ (or 35.4 THz), determined by the pump spectral width. This results in a spectral tuning speed of 5.6 THz/μs, far exceeding that of other advanced frequency-agile sources (e.g., 267 GHz/μs reported in the near-IR [12]). As detailed in Fig. 3(b), the chemical fingerprints are encoded on the pulses. Each pulse corresponds to a spectral element or point of the mid-IR spectrum. A time-encoded pulse sequence contains over 380 pulses or spectral points, resulting in a spectral point spacing of ~3.1 cm$^{-1}$. This is consistent with the result (3.08 cm$^{-1}$) derived from the relationship between $f_D$ and $\tau$. Specifically, the asynchronous temporal interval ($\Delta f_r/f_r^2$) is 0.77 ps, and $\beta$ = 0.12 THz/ps or 4 cm$^{-1}$/ps. Note that when using our tunable source for spectral measurements, the spectral resolution is set by the upper bound of the point spacing and the pulse spectral width (e.g., ~8 cm$^{-1}$).

To validate our spectral results, we convert the temporal data into the spectral domain (in wavenumbers) using the calculated spectral point spacing (3.08 cm$^{-1}$). Additionally, we calibrate the absolute wavenumber axis in advance using known absorption line centers of neat benzene. By normalizing the peaks of the mid-IR pulses against the background, we reconstruct the absorption spectrum of the chemical mixture. Figure 3(c) shows a portion of the normalized spectral data (red dots) on the calibrated wavenumber axis. Our results, which reveal the absorption bands of benzene (e.g., around 3072 cm$^{-1}$) and ethanol (around 2900 cm$^{-1}$), align well with the absorption spectrum (black curve in Fig. 3(c)) measured by the FTIR spectrometer (spectral resolution of 8 cm$^{-1}$). Notably, our data are obtained in a single-shot measurement of ~6.3 µs, whereas the FTIR measurement takes ~200 ms.

Moreover, data averaging improves the signal-to-noise ratio (SNR) of a spectral element, defined as the ratio between the pulse peak and the standard deviation (SD) of the noise floor. The SNR averaged over the spectral FWHM in a single-shot measurement is ~100, which improves to 3200 with 1000-fold averaging (Fig. 3(d)). The fitting line in Fig. 3(d) indicates that the SNR evolves as the square root of the number of averages.

The broad spectral coverage allows our approach to interrogate molecular fingerprints of various organic and biological samples. For instance, the reconstructed absorption spectra of a dimethyl sulfoxide (DMSO) solution (Fig. 4(a)) and an ethanol solution (Fig. 4(b)), with volume fractions of approximately 0.5% and 1%, respectively, diluted in carbon tetrachloride, are presented in the range of 2800 cm$^{-1}$ to 3200 cm$^{-1}$. The spectral results generally agree with the data obtained from the FTIR spectrometer. In addition, we measure absorption spectra of biological samples, such as flaxseed oil (Fig. 4(c)), in the spectral range from 2600 to 3780 cm$^{-1}$. The margin of error in Fig. 4(c) represents the fluctuations from 200 consecutive measurements. The FTIR data (black curve) are displayed for comparison.

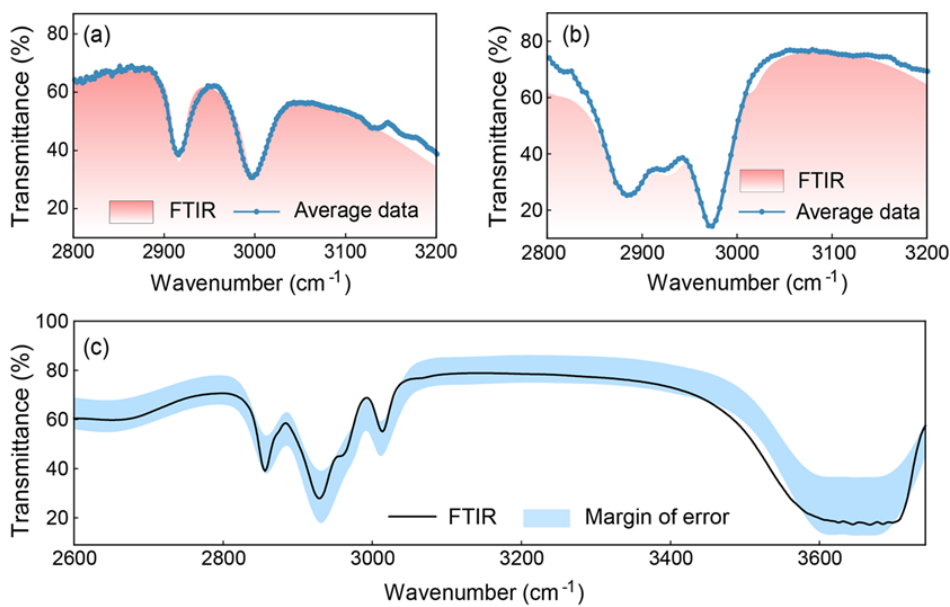

Fig. 4. Broadband spectral results. The reconstructed spectra of (a) dimethyl sulfoxide (DMSO) and (b) ethanol with 100-fold averaging. (c) Broadband absorption spectrum of flaxseed oil. In (c), the black curve is measured by a Fourier-transform infrared spectrometer (FTIR) at spectral resolution of 8 cm$^{-1}$.

*3.3 Spectral tuning at high scan rates*

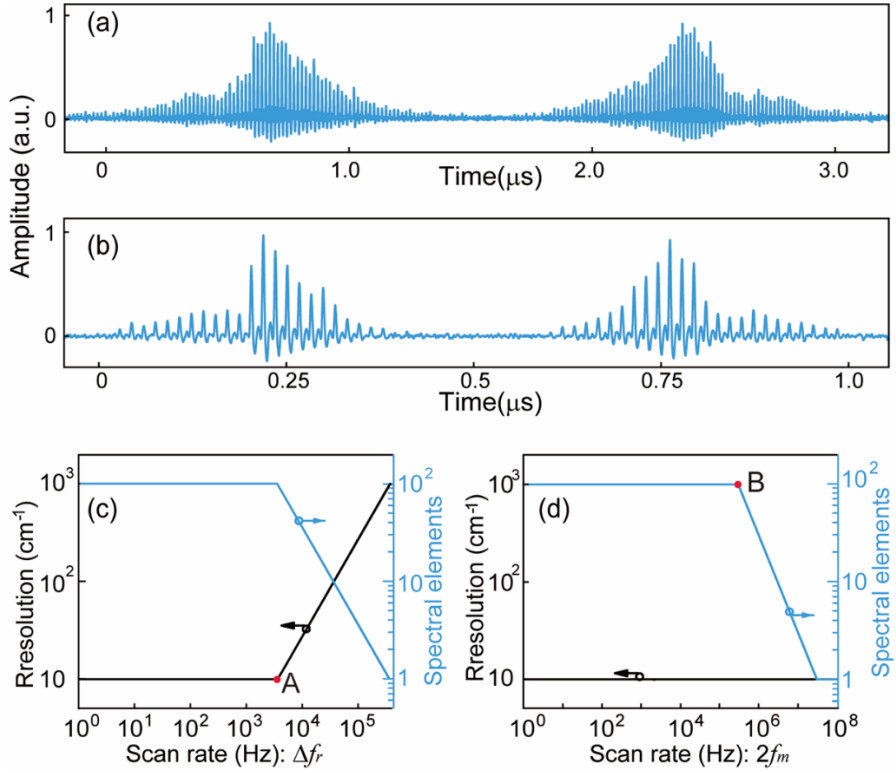

Fig. 5. Spectral tuning at high scan rates. Mid-IR pulse traces are recorded at (a) $2f_m$=600 kHz and $\Delta f_r$=60 kHz and (b) $2f_m$=2 MHz and $\Delta f_r$=200 kHz. A comparison of spectral parameters between (c) unmodulated and (d) frequency-modulated asynchronous schemes. For the plots in (c) and (d), we assume a full spectral span of 1000 cm$^{-1}$ and effective time duration of 100 ps. The spectral width of a single pulse is fixed at 10 cm$^{-1}$. The repetition frequencies of the fiber laser and the EO comb are $f_r$=60.5 MHz and $5f_r + 5\Delta f_r$=302.5 MHz + $5\Delta f_r$. Specifically, in (d), $\Delta f_r$ is fixed at 18 kHz.

A key merit of our source lies in the frequency-modulated EO comb, which enables high scan rates for tuning mid-infrared wavelengths. For instance, spectral tuning across ~1000 cm$^{-1}$ at a scan rate of $2f_m$=600 kHz, i.e., 600 kscans/s, is shown in Fig. 5(a). The scan rate can be further increased to 2 Mscans/s with $2f_m$=2 MHz, as shown in Fig. 5(b), at the cost of reduced pulse number and, consequently, an enlarged spectral point spacing. For example, there are ~100 pulses, spaced by 10 cm$^{-1}$, for one scan in Fig. 5(a), and ~30 pulses, spaced by 33 cm$^{-1}$, in Fig. 5(b). Nonetheless, the maximum tuning speed achievable with our setup reaches ~2000 cm$^{-1}$/μs or ~60 THz/μs, which is higher than that of a time-stretch mid-IR source (e.g., 19 cm$^{-1}$ at 50 MScans/s, or ~30 THz/μs [26]). These capabilities make our mid-IR source highly promising for multiplexed spectral interrogation of chemical dynamics on the μs scale [36]. Additionally, our source has immediate applications in multiplexed spectral imaging [16], optical coherence tomography [37], and parallel range measurements [3].

In particular, for spectral measurements, the frequency-modulated method we used offers significant advantages over conventional asynchronous schemes. In the conventional case (Fig. 5(c)), the scan rate (set by $\Delta f_r$) is directly linked with the spectral point spacing. When the spacing exceeds the pulse spectral width (e.g., indicated by point A in Fig. 5(c)), the continuous increase of $\Delta f_r$ degrades spectral resolution (black line), while simultaneously reducing the effective number (N) of spectral elements (blue line). Here, N is defined as a ratio between a full spectral width and spectral resolution. When N=1, only one spectral element is measured.

In contrast, in our case, the scan rate is determined by $2f_m$, which is independent of spectral resolution, as indicated by the black line in Fig. 5(d). This means that a high scan rate can be achieved without sacrificing spectral resolution, provided other parameters (such as $\Delta f_r$ and $\beta$) remain unchanged. However, when the scan rate reaches a point (point B on the blue line in Fig. 5(d)) where the scan duration equals to the effective time window (see Fig. 1(d)), any further increase in the scan rate narrows the spectral tuning range, causing a reduction in N. Nonetheless, for spectroscopic applications, the frequency-modulated scheme can provide a scan rate that is several orders of magnitude higher than that of the unmodulated case, while preserving the spectral resolution and the number of spectral elements.

Currently, the finest spectral resolution offered by our mid-IR source is about 8 cm$^{-1}$, limited by the spectral focusing process. This process can be improved through refined dispersion management of the pump and signal pulses, including compensation of high-order dispersion [38]. Nevertheless, this resolution is sufficient for characterizing condensed samples, which typically exhibit broad absorption bands rather than sharp absorption lines, as seen in gas-phase samples.

## 4. Discussion

Our tunable mid-IR source and its application in spectroscopy share a similar architecture to dual-comb spectroscopy, utilizing two asynchronous lasers and a photodetector. Dual-comb spectroscopy is a transformative technique that offers unparalleled spectral resolution, precision, and accuracy for rapid broadband molecular spectroscopy [4-10]. However, it requires two mutually coherent combs, which presents technical challenges [6] and adds complexity to the system. Furthermore, achieving broadband (>1000 cm$^{-1}$) mid-IR spectra within a few microseconds remains a significant challenge for dual-comb spectrometers, primarily due to Nyquist limitations [5]. In contrast, our approach circumvents the need for coherent comb generation and phase stabilization, enabling wavelength-tunable spectral detection directly in the time domain without relying on Fourier transformation.

Another technique with similarities to our approach is photonic time-stretching. In this method, a laser pulse is temporally dispersed, establishing an intra-pulse time-wavelength correspondence for ultrafast spectral analysis and signal processing [39]. Recent demonstrations of mid-IR time-stretch lasers for spectroscopy have enabled single-shot spectral measurements on the nanosecond timescale, with scan rates at the pulse repetition frequency (typically tens of MHz [26, 36, 40, 41]). However, the pulse energy is dispersed, limiting the SNR per spectral elements (e.g., a peak SNR of 14 reported in [26]), and the spectral width in these systems is typically restricted to tens of cm$^{-1}$. Furthermore, these systems face significant challenges in high-speed mid-IR detection (e.g., bandwidth of >10 GHz [36]) as well as data storage and processing. In contrast, our source enables broadband mid-IR spectral tuning (1180 cm$^{-1}$) on a pulse-to-pulse basis, offering a high SNR (~100) while alleviating the burdens on detection and digitization, as evidenced by our use of a 250 MHz detection bandwidth and a moderate sampling rate (500 MS/s).

Finally, with recent advancements in nanophotonics, particularly the development of on-chip electro-optical combs [42], our approach holds the potential to lead to compact, wavelength-swept laser sources with broad applications across chemistry, biology, and materials science.

## 5. Summary

In summary, we present a method for ultra-rapid, broadband wavelength sweeping in the mid-IR region, utilizing spectral focusing DFG between a broadband fiber laser and an asynchronous, frequency-modulated EO comb. This approach bypasses mechanical scanning limitations, enabling automatic spectral tuning at 5.6 THz/μs and pulse-to-pulse spectral encoding. Applied to spectroscopic sensing, it offers broad spectral detection (1180 cm$^{-1}$ or 35.4 THz) with moderate resolution (8 cm$^{-1}$) and rapid measurement times (~6.3 μs). The cavity-free EO comb facilitates fast repetition frequency modulation, achieving high scan rates (up to 2 MHz, corresponding to a tuning speed of ~60 THz/μs), albeit with reduced resolution (33 cm$^{-1}$). Despite this, our method shows promise for broadband spectral sensing of condensed-phase samples and ultra-rapid mid-IR wavelength-space encoding, with potential applications in chemical monitoring, mid-IR flow cytometry, biomedical imaging, and scanner-free LIDAR.

**Funding.** Innovation Program for Quantum Science and Technology (2023ZD0301000).